%% file: main.tex
\documentclass[journal]{IEEEtran}
\usepackage{amsmath,amssymb,amsfonts}
\usepackage[caption=false,font=normalsize,labelfont=sf,textfont=sf]{subfig}
\usepackage{amsthm}
\usepackage{algorithmic}
\usepackage{algorithm}
\usepackage{graphicx}
\usepackage{tcolorbox}
\usepackage{textcomp}
\usepackage{xcolor}
\usepackage{xspace}
\usepackage{cite}
\usepackage{hyperref}
\usepackage{stfloats}
\usepackage{url}
\usepackage{verbatim}
\usepackage{booktabs} 
\usepackage{multirow} 
\usepackage{bm}

\usetikzlibrary{positioning,fit,calc,backgrounds,shapes.symbols}
\usetikzlibrary{
	shapes.geometric,
	shapes.symbols,
	arrows.meta,
	positioning,
	backgrounds,
	fit
}

\title{Chip Floorplanning Combining Convex and Non-convex Optimization}

\usepackage{orcidlink}
\author{
Yilihamujiang Yimamu\,\orcidlink{0000-0002-1068-8625},
Guillem Pastor-Rué\,\orcidlink{0009-0001-3836-2475}, 
and Jordi Cortadella\,\orcidlink{0000-0001-8114-250X}%\IEEEmembership{Fellow, IEEE}
\thanks{The authors are with the Department of Computer Science, Universitat Politècnica de Catalunya, Barcelona, Spain.}
\thanks{This work has been supported by the Càtedra Chip UPC project, funded by the Spanish Ministry (Ministerio para la Transformación Digital y de la Función Pública) and the European Union – Next Generation EU (aid file TSI-069100-2023-0015).
}}

\begin{document}
\input{definitions}
\maketitle

\input{Abstract}
\input{Introduction}
\input{Preliminaries}

\input{GlobalFloorplan}

\input{Numericaltests}

\input{Conclusion}
\input{Acknowledgment}

\input{Appendix}

\bibliographystyle{ieeetr}
\bibliography{main}

\end{document}

%% file: definitions.tex
% Definitions for the graph
\newcommand{\vertices}{\ensuremath{\mathcal{V}}}
\newcommand{\edges}{\ensuremath{\mathcal{E}}}
\newcommand{\netlist}{\ensuremath{\mathcal{N}}}
\newcommand{\xnetlist}{\netlist{\vertices,\edges}}

% Definitions for the rectangles
\newcommand{\coord}{\ensuremath{\mathbf{c}}\xspace}
\newcommand{\rdim}{\ensuremath{\textbf{shp}}\xspace}  %\dim already defined
\newcommand{\ar}{\ensuremath{AR}} 

% Wirelength models
\newcommand{\wl}{\text{WL}\xspace}
\newcommand{\wlhpwl}{\ensuremath{\wl}_{\text{HPWL}}}
\newcommand{\wllse}{\ensuremath{\wl}_{\text{LSE}}}
\newcommand{\wlquad}{\ensuremath{\wl}_{\text{Quad}}}
\newcommand{\sspan}[2]{\ensuremath{\widehat{\text{Span}_#1}(#2)}} % \span already defined

% Die
\newcommand{\die}{\textbf{D}\xspace}
\newcommand{\dwidth}{\ensuremath{\die_\text{W}}}
\newcommand{\dheight}{\ensuremath{\die_\text{H}}}
\newcommand{\dlength}{\ensuremath{\die_\text{L}}}

%Prelim
\newcommand{\refarea}{\ensuremath{I_{ij}}}
\newcommand{\cdist}{\ensuremath{d}}
\newcommand{\mcdist}{\ensuremath{\bar \cdist}}
\newcommand{\pdepth}{\ensuremath{p}}
\newcommand{\ratio}{\ensuremath{R}}
\newcommand{\olp}{\ensuremath{O}}

%% file: Abstract.tex
\begin{abstract}
Floorplanning is a critical early stage of VLSI physical design, and its quality directly impacts interconnect wirelength, chip performance, and downstream design efficiency. This article presents a multi-stage fixed-outline floorplanning framework that combines non-convex and convex optimization. The framework operates in three successive phases. In the initial floorplanning phase, a quadratic placement optimizes module connectivity to produce a topology-aware starting configuration with strong net clustering, albeit with significant module overlaps. In the global floorplanning phase, Adam-based projected gradient method is employed to solve non-convex optimization that minimizes wirelength and overlap. In the legalization phase, a log-transformed model exploiting horizontal and vertical constraint graphs eliminates residual overlaps under a convex formulation. Experiments on the MCNC, GSRC and HB+ benchmarks demonstrate that the framework achieves state-of-the-art wirelength quality, reducing average HPWL by at least 1\% and 5\% compared to state-of-the-art floorplanners on tested benchmarks. 
\end{abstract}


%% file: Introduction.tex
\section{Introduction}

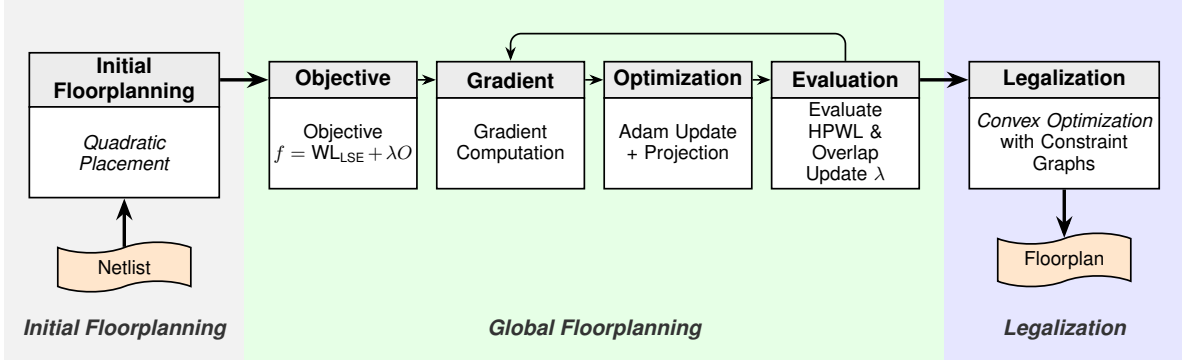
\begin{figure*}%[h]
    \centering
    \scalebox{0.8}{\input{pict/flow_tikz}}
    \caption{Floorplanning flow}
    \label{fig:flow}
\end{figure*}

\IEEEPARstart{F}{loorplanning} is one of the early stages of physical design.
Traditionally, this task heavily relied on human intervention and the skills of expert designers.
Today, large chips contain hundreds of blocks, and floorplanning is becoming a challenging problem that demands automation. Floorplanning is a combinatorial optimization problem that can be tackled at different levels of abstraction, depending on the mathematical models and algorithms used for optimization. Fixed-outline floorplanning fundamentally requires balancing two competing objectives: reducing interconnect wirelength while simultaneously enforcing strict geometric feasibility. 

Unfortunately, even simplified variants of this task, such as rectangle packing, are known to be NP-complete~\cite{Korf2003}. Finding optimal solutions is generally impractical for realistic problem sizes, and most approaches rely on approximation strategies to explore a highly nonconvex solution space. When mathematical optimization frameworks are adopted, further challenges emerge. Wirelength must be approximated using tractable surrogate models, and the resulting convex or nonconvex optimization landscape makes the solution process highly sensitive to initialization. In practice, starting from a high-quality initial configuration is often essential for avoiding poor local minima and achieving competitive results~\cite{Ren2025}.

\subsection{Previous work}
Decades of research into fixed-outline floorplanning have produced a diverse landscape of solutions, ranging from classical discrete heuristics and multilevel frameworks to analytical optimization and emerging learning-based models.

Early floorplanning approaches primarily relied on discrete representations combined with meta-heuristic search strategies. Representative floorplan representations include O-tree \cite{guo1999tree}, B*-tree \cite{chang2000b}, and Sequence Pairs \cite{murata1995rectangle}. These representations encode the topological relations among modules, after which simulated annealing (SA) is commonly employed to explore the solution space. For example, Parquet \cite{adya2004fixed} adopts sequence pair with SA-based optimization. While such discrete meta-heuristic methods are effective for small-scale problems, their runtime grows rapidly with problem size due to the exponential search space. Moreover, as more practical constraints are incorporated, the accumulation of complex heuristics often leads to unpredictable tool behavior and limited scalability.

To address scalability issues, multilevel heuristic frameworks have been proposed. 
The core idea is to reduce the search space progressively through clustering or partitioning, thereby enabling large-scale floorplanning. Clustering-based approaches, such as MB*-tree \cite{lee2003multilevel}, merge modules hierarchically until the reduced problem becomes manageable. The solution is then refined during successive declustering stages. However, early clustering decisions are made without a global perspective, potentially limiting final solution quality.

Partitioning-based approaches decompose the original design into smaller subproblems, which are solved individually and then merged level by level. Representative methods include Capo \cite{adya2004unification}, DeFer \cite{yan2010defer}, IMF \cite{chen_new_2008}, and QinFer \cite{ji_quasi-newton-based_2021}, many of which rely on hMetis \cite{karypis1999multilevel} for hypergraph partitioning. Compared with clustering-based methods, partitioning-based approaches can incorporate wirelength information earlier in the process. Nevertheless, accurate wirelength estimation during partitioning remains challenging, and solution quality may degrade due to approximation errors introduced in early decomposition.

Inspired by analytical placement techniques, several analytical floorplanning methods formulate the problem as a constrained optimization model. Typical frameworks consist of two major stages: global floorplanning, which determines rough module positions, and legalization, which removes overlaps and determines exact dimensions of soft modules.

During the global floorplanning phase, pioneering approaches such as AR \cite{luo_large-scale_2008} and UFO \cite{lin2011ufo} simplify the problem by modeling modules as basic geometric shapes (e.g., squares or circles), applying force-directed mechanisms to drive the optimization. To provide theoretical guarantees on solution quality, semidefinite programming (SDP)-based \cite{li_global_2023} formulations have been introduced, albeit at a higher complexity. A significant shift in the field has been the adoption of density-based paradigms. For instance, analytical \cite{zhan_fixed-floorplanning_2006}, F-FM \cite{lin2014f} and Lin \textit{et al.} \cite{lin2018fast} adapt placement-centric density functions to handle soft modules, ensuring a uniform distribution across the chip area. Building upon this, PeF \cite{li_pef_2023} integrates floorplanning into an electrostatic optimization engine. 

The transition from global floorplanning to a legal layout remains a critical challenge. While second-order cone programming (SOCP)-based legalization offers mathematical rigor, its scalability is limited when handling moderate to large-scale instances. Consequently, many contemporary methods resort to more scalable alternatives; while some employ tree-based representations or integer linear programming (ILP), others, like PeF, alternatively modified relative positions of modules based on horizontal and vertical constraint graphs to eliminate overlaps.

More recently, learning-based methods have been explored for floorplanning. 
These approaches leverage reinforcement learning \cite{yang_miracle_2024} or graph neural networks \cite{liu_graphplanner_2022} to extract structural features from netlists and guide placement decisions. While promising in representation learning and search guidance, existing learning-based methods primarily focus on netlist encoding and objective prediction, with limited exploration of multi-constraint handling and legalization robustness in large-scale industrial settings.

Despite the continuous evolution of analytical floorplanning, achieving high-quality layouts for modern complex chips remains a formidable challenge, primarily driven by three interrelated issues. First, the inherent non-convexity of the floorplanning optimization landscape makes gradient-based solvers highly sensitive to the initial configuration, often trapping the optimization in suboptimal local minima. This sensitivity is further exacerbated by the extreme size disparity in mixed-size designs, where the objective function is dominated by large macros. Such an imbalance often leads to the 'drowning' of small modules, which are passively displaced into clusters of massive overlap, severely compromising the potential for wirelength reduction. Consequently, these flawed global distributions impose an excessive burden on the legalization stage. 

\subsection{A new multi-stage floorplanning framework}

The previous observations motivate the development of a unified, multi-stage framework that synergistically integrates gradient-based non-convex optimization for global floorplanning and convex optimization for legalization. 

To this end, this paper introduces an approach that progresses through successive levels of abstraction. It starts with a point-based quadratic placement to establish a topology-aware initial configuration. This is followed by a non-convex analytical model to globally optimize module locations and minimize overlaps. The process culminates in a convex reformulation to eliminate all residual overlaps and establish strict geometric feasibility, achieving a high-quality legal solution with minimized wirelength. The primary contributions of this work are summarized as follows:
\begin{enumerate}
    \item We design an analytical global floorplanning stage driven by an Adam-based projected gradient method. Instead of relying on discrete heuristics, we formulate a continuously differentiable non-convex objective that simultaneously minimizes the wirelength and a normalized smooth overlap penalty. The Adam optimizer adaptively navigates the optimization landscape, while projection operations strictly enforce die boundaries and module aspect-ratio constraints.
    \item We develop a robust legalization methodology to reconcile geometric feasibility with interconnect optimality. To clear the residual overlaps left by the global stage, we construct a mathematically rigorous convex optimization model. By leveraging a hybrid variable space with logarithmic transformations and constraint graphs, this approach translates the inherently non-convex legalization constraints into a tractable convex program, ensuring global optimality for the established topology.
\end{enumerate}

%% file: pict/flow_tikz.tex
\begin{tikzpicture}[
    font=\sffamily\small,
    >=Stealth,
    % --- 尺寸配置区 (可根据页面宽度自由调整) ---
    declare function={
        MainBoxWidth=3.0cm;
        SubBoxWidth=2.3cm;
    },
    % --- 样式定义区 ---
    base box/.style={
        align=center, 
        inner xsep=2pt, 
        inner ysep=2pt
    },
    block/.style={
        base box, 
        draw, thick, 
        fill=white,
        minimum height=1.5cm
    },
    header/.style={
        base box, 
        draw, thick, 
        fill=gray!15, % 稍微柔和的表头背景色
        minimum height=0.6cm, 
        font=\sffamily\bfseries
    },
    main box/.style={text width=MainBoxWidth},
    sub box/.style={text width=SubBoxWidth},
    data/.style={
        draw, thick, 
        fill=orange!20, 
        shape=tape,
        minimum width=2.2cm, 
        minimum height=0.7cm,
        font=\sffamily\small
    },
    stage label/.style={
        font=\sffamily\bfseries\itshape,
        text=black!80
    }
]

%%%%%%%%%%%%%%%%%%%%%%%%%%%%%%%%%%%%%%%%%%%%%%%%%%%%%%
%% 1. Initial Floorplanning
%%%%%%%%%%%%%%%%%%%%%%%%%%%%%%%%%%%%%%%%%%%%%%%%%%%%%%
\node[header, main box] (ip_head) {Initial Floorplanning};
% 使用 yshift=0.8pt 完美抵消线条宽度，实现无缝贴合
\node[block, main box, anchor=north, yshift=0.8pt] (ip_desc) at (ip_head.south) {\textit{Quadratic}\\\textit{Placement}};

\node[data, below=0.8cm of ip_desc] (netlist) {Netlist};

%%%%%%%%%%%%%%%%%%%%%%%%%%%%%%%%%%%%%%%%%%%%%%%%%%%%%%
%% 2. Global Floorplanning
%%%%%%%%%%%%%%%%%%%%%%%%%%%%%%%%%%%%%%%%%%%%%%%%%%%%%%
\node[header, sub box, right=0.8cm of ip_head] (obj_head) {Objective};
\node[block, sub box, anchor=north, yshift=0.8pt] (obj_desc) at (obj_head.south) {Objective\\$f=\wllse+\lambda O$};

\node[header, sub box, right=0.3cm of obj_head] (grad_head) {Gradient};
\node[block, sub box, anchor=north, yshift=0.8pt] (grad_desc) at (grad_head.south) {Gradient\\Computation};

\node[header, sub box, right=0.3cm of grad_head] (adam_head) {Optimization};
\node[block, sub box, anchor=north, yshift=0.8pt] (adam_desc) at (adam_head.south) {Adam Update\\+ Projection};

\node[header, sub box, right=0.3cm of adam_head] (eval_head) {Evaluation};
\node[block, sub box, anchor=north, yshift=0.8pt] (eval_desc) at (eval_head.south) {Evaluate\\HPWL \& Overlap\\Update $\lambda$};

%%%%%%%%%%%%%%%%%%%%%%%%%%%%%%%%%%%%%%%%%%%%%%%%%%%%%%
%% 3. Legalization
%%%%%%%%%%%%%%%%%%%%%%%%%%%%%%%%%%%%%%%%%%%%%%%%%%%%%%
\node[header, main box, right=0.8cm of eval_head] (conv_head) {Legalization};
\node[block, main box, anchor=north, yshift=0.8pt] (conv_desc) at (conv_head.south) {\textit{Convex Optimization}\\with Constraint Graphs};

\node[data, below=0.8cm of conv_desc] (floorplan) {Floorplan};

%%%%%%%%%%%%%%%%%%%%%%%%%%%%%%%%%%%%%%%%%%%%%%%%%%%%%%
%% Connections 
%%%%%%%%%%%%%%%%%%%%%%%%%%%%%%%%%%%%%%%%%%%%%%%%%%%%%%
% 
\draw[->, ultra thick] (netlist.north) -- (ip_desc.south);
\draw[->, ultra thick] (conv_desc.south) -- (floorplan.north);

\draw[->, ultra thick] (ip_head.east) -- (obj_head.west);
\draw[->, thick]       (obj_head.east) -- (grad_head.west);
\draw[->, thick]       (grad_head.east) -- (adam_head.west);
\draw[->, thick]       (adam_head.east) -- (eval_head.west);
\draw[->, ultra thick] (eval_head.east) -- (conv_head.west);

\draw[->, thick, rounded corners=5pt] 
    (eval_head.north) -- ++(0, 0.45) -| (grad_head.north);

%%%%%%%%%%%%%%%%%%%%%%%%%%%%%%%%%%%%%%%%%%%%%%%%%%%%%%
%% Background Colors & Bottom Stage Labels
%%%%%%%%%%%%%%%%%%%%%%%%%%%%%%%%%%%%%%%%%%%%%%%%%%%%%%

\coordinate (bgTop)   at ($(ip_head.north) + (0, 0.85cm)$); 
\coordinate (LabelLine) at ($(netlist.south) - (0, 0.7cm)$);
\coordinate (bgBottom)  at ($(LabelLine) - (0, 0.5cm)$); % 
\coordinate (bgLeft)  at ($(ip_desc.west) - (0.4cm, 0)$);
\coordinate (cutA)    at ($(ip_desc.east)!0.5!(obj_desc.west)$);
\coordinate (cutB)    at ($(eval_desc.east)!0.5!(conv_desc.west)$);
\coordinate (bgRight) at ($(conv_desc.east) + (0.4cm, 0)$);

\begin{scope}[on background layer]
    \fill[gray!10]  (bgLeft |- bgTop) rectangle (cutA |- bgBottom);
    \fill[green!10] (cutA |- bgTop)   rectangle (cutB |- bgBottom);
    \fill[blue!10]  (cutB |- bgTop)   rectangle (bgRight |- bgBottom);
\end{scope}

\node[stage label] at (ip_desc |- LabelLine) {Initial Floorplanning};

\coordinate (global_center) at ($(obj_desc.west)!0.5!(eval_desc.east)$);
\node[stage label] at (global_center |- LabelLine) {Global Floorplanning};

\node[stage label] at (conv_desc |- LabelLine) {Legalization};

\end{tikzpicture}

%% file: Preliminaries.tex
\section{Preliminaries}
This section introduces the mathematical models for the fixed-outline floorplanning problem. We will present a non-convex optimization model and detail the critical challenge of incorporating non-overlap constraints.

\subsection{Fixed-outline Floorplanning Problem}

The fixed-outline region $\die=(\dwidth,\dheight)$ is defined as a rectangle, where $\dwidth$ and $\dheight$ represent the width and height of the die, the netlist can be modeled as a hypergraph $\mathcal{G}=(\vertices,\edges)$, where $\vertices:=\{v_1,v_2,...,v_n\}$ represents the set of modules and $\edges:\{e_1,e_2,...,e_m\}$ denotes the set of nets with weights $\omega_{e_k}$. Every module $v_i\in \mathcal{V}$ is modeled as a rectangle parameterized by its center coordinates $\coord_i=(x_i,y_i)$ and its shapes $\rdim_i=(w_i, h_i)$,
 the fixed-outline floorplanning problem can be formulated as follows:
%\begin{eqnarray}\label{eq1}
%\min&&\wl(\mathcal{E})\\
%s.t.&& \text{modules do not overlap}\nonumber\\
%  &&\text{modules are bounded in the fixed outline}\nonumber\\
%  && \text{suitable aspect ratio of the modules} \nonumber\\
%  && \text{modules meet target area} \nonumber
%\end{eqnarray}

\begin{eqnarray}\label{eq1}
\min&&\wl(\mathcal{E})\\
s.t.&& \text{modules} \left\{
\begin{array}{l}
\text{do not overlap}\nonumber \\
\text{meet target area} \nonumber \\
\text{meet aspect ratio} \nonumber\\
\text{are inside the die} \nonumber
\end{array}
\right.
\end{eqnarray}
where \wl represents a wirelength model for the nets.

\subsection{Wirelength models}\label{Wl}
The Half-Perimeter Wirelength (HPWL) model is the most common metric for estimating the length of a net $e_k\in \mathcal{E}$ connecting a set of modules:
$$\wlhpwl(e_k): = \omega_{e_k}\left( \max_{i\in e_k}x_i-\min_{i\in e_k}x_i+\max_{i\in e_k}y_i -\min_{i\in e_k}y_i\right).$$
However, the $\max$ and $\min$ functions are non-differentiable, making it unsuitable for conventional gradient-based optimization algorithms like those used in analytical floorplanner. 

To overcome this, one common approximation models the wirelength as the sum of weighted squared Euclidean distances between module centers. This can be integrated into a quadratic objective function:
$$\wlquad :=\frac{1}{2}\sum_{i,j}\omega_{ij}((x_i-x_j)^2+(y_i-y_j)^2),$$
where the weight $\omega_{ij}$ represents the connectivity between modules $i$ and $j$. Under the standard clique model approximation for multi-terminal nets, a net $e$ containing $\vert{}e\vert{}$ pins is modeled as a fully connected graph (clique). For each pair of modules $i, j \in e$, a uniform weight is assigned to the edge connecting them. The total weight $\omega_{ij}$ is the sum of contributions from all nets shared by both modules:
$$\omega_{ij} = \sum_{e_k \in E_{ij}} \frac{2 \cdot \omega_{e_k}}{\vert{}e_k\vert{}(\vert{}e_k\vert{}-1)},$$
where $E_{ij}$ is the set of all nets that connect both module $i$ and module $j$. This scaling factor ensures that the total weight of the clique graph equals the original weight of the net, distributing the wirelength penalty evenly among its components.

To achieve a higher fidelity to the HPWL while maintaining differentiability, the Log-Sum-Exp (LSE) model is often employed. LSE is a popular smooth approximation where the $x$-direction span of the bounding box for a net is defined as:
$$\sspan{x}{e_k}=\alpha\left ( \log\left ( \sum_{i\in e_k}e^{x_i/\alpha} \right )+ \log\left ( \sum_{i\in e_k}e^{-x_i/\alpha} \right )  \right ),$$
where $\alpha>0$ is a fixed smoothing parameter during optimization. A smaller $\alpha$ yields a closer approximation to the true maximum but makes the function steeper. The total wirelength for a net is then calculated as:
\begin{align}\label{LSE}
\wllse(e_k)=\sspan{x}{e_k}+\sspan{y}{e_k}.    
\end{align}
This model is widely used in state-of-the-art analytical placers because it balances mathematical smoothness with a very tight fit to the actual HPWL.

\subsection{Module overlapping}\label{non-overlap}

To reduce module overlap while preserving the differentiability required by gradient-based optimization, we introduce a normalized smooth overlap penalty. For two modules $i$ and $j$ centered at $(x_i,y_i)$ and $(x_j,y_j)$ with widths $w_i,w_j$ and heights $h_i,h_j$, we first define the center distances along the two coordinate axes as

\begin{equation}
\cdist_{ij}^{x}=\sqrt{(x_i-x_j)^2+\delta},\qquad
\cdist_{ij}^{y}=\sqrt{(y_i-y_j)^2+\delta}.
\end{equation}
where $\delta$ is a small positive constant for numerical stability. 
The minimum center distances required to avoid overlap are

\begin{equation}
\mcdist_{ij}^{x}=\frac{w_i+w_j}{2},\qquad
\mcdist_{ij}^{y}=\frac{h_i+h_j}{2}.
\end{equation}

Instead of directly measuring the overlap length, we normalize the penetration depth by the corresponding non-overlapping distance

\begin{equation}
\pdepth_{ij}^{x}
=
\frac{\mcdist_{ij}^{x}-\cdist_{ij}^{x}}
{\mcdist_{ij}^{x}},
\qquad
\pdepth_{ij}^{y}
=
\frac{\mcdist_{ij}^{y}-\cdist_{ij}^{y}}
{\mcdist_{ij}^{y}},
\end{equation}
This normalization prevents large modules from producing disproportionately large gradients and therefore improves optimization robustness.

Since the overlap penalty should vanish when two modules are separated while remaining differentiable everywhere, we approximate the positive operator using the Softplus function

\begin{equation}
r_{ij}^{x}
=
\tau
\log
\left(
1+
\exp
\left(
\frac{\pdepth_{ij}^{x}}{\tau}
\right)
\right),
\end{equation}

\begin{equation}
r_{ij}^{y}
=
\tau
\log
\left(
1+
\exp
\left(
\frac{\pdepth_{ij}^{y}}{\tau}
\right)
\right),
\end{equation}
where $\tau$ controls the smoothness of the approximation. The two-dimensional normalized overlap ratio is then defined as

\begin{equation}
\ratio_{ij}=r_{ij}^{x}r_{ij}^{y}.
\end{equation}

To convert the normalized overlap ratio into a physical penalty, we introduce the reference area

\begin{equation}
\refarea
=
\mcdist_{ij}^{x}\mcdist_{ij}^{y},
\end{equation}
which corresponds to the minimum bounding region where the two modules start to interfere. The pairwise overlap cost is formulated as

\begin{equation}
\olp_{ij}
=
\ratio_{ij}^{2}\refarea.
\end{equation}

The quadratic formulation increases the penalty for severe overlaps while maintaining smooth gradients for small violations. Finally, the total overlap objective is obtained by summing all valid module pairs,

\begin{equation}
\mathcal{L}_{\mathrm{ov}}
=
\frac12
\sum_{i=1}^{N}
\sum_{j=1}^{N}\olp_{ij},
\end{equation}
where the factor $\frac12$ avoids double counting of symmetric pairs.

%% file: GlobalFloorplan.tex
\section{Floorplanning Algorithm}

This section presents the initial, global floorplanning and legalization procedures of the framework. 

The initial location of the modules is pre-computed using quadratic placement~\cite{Alpert97QP}. The locations are calculated by solving a sparse system of linear equations, minimizing the $\wlquad$, and assuming fixed terminals at the boundaries of the die. The transformation from hypergraph to graph is done using the clique model for hyperedges. Rather than starting from a random placement or using a force-directed initialization, we employ QP because it provides a high-quality starting point by preserving the topological structure of the netlist in the 2D embedding while producing a low-wirelength placement. Starting from this structured initialization significantly improves the effectiveness of the subsequent optimization.

\subsection{Global Floorplanning Algorithm}

The objective of global floorplanning is to determine the locations of soft modules that minimize the total HPWL while allowing a limited amount of overlap. This task is formulated as an optimization problem by incorporating the overlap constraint into the objective using a penalty coefficient $\lambda$:
\begin{eqnarray}\label{eq_gf}
\min&&\sum_{e_k \in \mathcal{E}} \wllse(e_k) + \lambda\cdot\mathcal{L}_{\mathrm{ov}}\\
s.t.&& \ar_{\min}\leq \frac{w_i}{h_i}\leq \ar_{\max}\nonumber\\
&& \text{all modules satisfy the fixed-outline constraint.}\nonumber
\end{eqnarray}
Here, two constraints are imposed on the modules during optimization. First, the aspect ratio of each soft module is constrained to lie within prescribed bounds $[\ar_{\min},\ar_{\max}]$. Second, all modules are constrained to remain within the fixed outline. In our formulation, the dimensions of soft modules are treated as decision variables while the dimensions of hard and fixed modules remain constants.

Unlike the global placement stage in analytical placers, which typically enforces a global density constraint to achieve uniform module distribution, global floorplanning focuses primarily on minimizing module overlap. Consequently, we can directly incorporate the overlap cost as a penalty term within our objective function. 

\subsubsection{Gradient Derivation}

Since the wirelength function $\wllse(e_k)$ is expressed in an analytical form, its partial derivatives are straightforward to compute. We therefore focus on the overlap objective, which is composed of elementary differentiable operations; its gradients with respect to module locations and dimensions are analytically derived using the chain rule as follows:

The derivative of the pairwise overlap cost is

\begin{equation}
\frac{\partial \olp_{ij}}{\partial \theta}
=
2\ratio_{ij}\refarea
\frac{\partial \ratio_{ij}}{\partial \theta}
+
\ratio_{ij}^{2}
\frac{\partial \refarea}{\partial \theta},
\end{equation}
where $\theta\in\{x_i,y_i,w_i,h_i\}$.

The derivatives of the normalized overlap ratio are

\begin{equation}
\frac{\partial \ratio_{ij}}{\partial \theta}
=
r_{ij}^{y}
\frac{\partial r_{ij}^{x}}{\partial \theta}
+
r_{ij}^{x}
\frac{\partial r_{ij}^{y}}{\partial \theta},
\end{equation}
and the Softplus function yields

\begin{equation}
\frac{\partial r}{\partial \pdepth}
=
\frac{1}{1+\exp(-\pdepth/\tau)}
\triangleq
\sigma\!\left(\frac{\pdepth}{\tau}\right),
\end{equation}
where $\sigma(\cdot)$ denotes the sigmoid function.

By substituting the intermediate geometric derivatives into the chain rule, we obtain the exact gradients of the overlap penalty with respect to the module center coordinates $(x_i, y_i)$ as
\begin{align}
\frac{\partial \olp_{ij}}{\partial x_i} &= -2\ratio_{ij}\refarea r_{ij}^{y} \, \sigma\!\left(\frac{\pdepth_{ij}^{x}}{\tau}\right) \frac{x_i-x_j}{\mcdist_{ij}^{x}\cdist_{ij}^{x}}, \label{eq:grad_x} \\
\frac{\partial \olp_{ij}}{\partial y_i} &= -2\ratio_{ij}\refarea r_{ij}^{x} \, \sigma\!\left(\frac{\pdepth_{ij}^{y}}{\tau}\right) \frac{y_i-y_j}{\mcdist_{ij}^{y}\cdist_{ij}^{y}}. \label{eq:grad_y}
\end{align}
Due to the symmetry of the distance function, the gradients with respect to $(x_j, y_j)$ can be directly obtained by negating the expressions in Eq. (\ref{eq:grad_x}) and Eq. (\ref{eq:grad_y}).

Similarly, the derivatives with respect to the module dimensions $(w_i, h_i)$ must account for the variations in both the normalized overlap depth and the reference bounding area $\refarea$. The closed-form gradients are derived as
\begin{align}
\frac{\partial \olp_{ij}}{\partial w_i} &= \ratio_{ij}\refarea r_{ij}^{y} \, \sigma\!\left(\frac{\pdepth_{ij}^{x}}{\tau}\right) \frac{\cdist_{ij}^{x}}{(\mcdist_{ij}^{x})^2} + \frac{1}{2}\ratio_{ij}^{2}\mcdist_{ij}^{y}, \\
\frac{\partial \olp_{ij}}{\partial h_i} &= \ratio_{ij}\refarea r_{ij}^{x} \, \sigma\!\left(\frac{\pdepth_{ij}^{y}}{\tau}\right) \frac{\cdist_{ij}^{y}}{(\mcdist_{ij}^{y})^2} + \frac{1}{2}\ratio_{ij}^{2}\mcdist_{ij}^{x}.
\end{align}
Since $\mcdist_{ij}^x$ and $\mcdist_{ij}^y$ are symmetric with respect to widths and heights, it naturally follows that $\frac{\partial \olp_{ij}}{\partial w_j} = \frac{\partial \olp_{ij}}{\partial w_i}$ and $\frac{\partial \olp_{ij}}{\partial h_j} = \frac{\partial \olp_{ij}}{\partial h_i}$. This fully analytical and continuously differentiable formulation enables highly efficient physical gradient computation on modern hardware accelerators during the end-to-end optimization process.

Unlike the overlap-area formulation \cite{chen2024modern}, whose gradients are determined by the active overlap boundaries through piecewise derivatives of the $\min(\cdot)$ and $\max(\cdot)$ operators, the proposed formulation is expressed solely by normalized center-distance violations. Consequently, the gradients admit a unified closed-form expression without explicitly enumerating geometric overlap cases. This property considerably simplifies the gradient computation while maintaining meaningful separation forces under all overlap configurations, including partial overlap and full containment.

In Algorithm \ref{alg:jax-glb-algic}, a projected gradient method is used for \ref{eq_gf}, while enforcing die-boundary, aspect-ratio constraints through projection after each update. First, the objective function $f_k$ required in the k-th iteration is constructed, where the initial penalty factor $\lambda_0$ is determined by gradient balancing, ensuring that the initial magnitudes of the wirelength and overlap gradients are comparable:
\begin{equation}
\lambda_0 = \frac{\sum_i \left( \left| \frac{\partial \wllse}{\partial x_i} \right| + \left| \frac{\partial \wllse}{\partial y_i} \right| \right)}{\sum_i \left( \left| \frac{\partial \olp_{ij}}{\partial x_i} \right| + \left| \frac{\partial \olp_{ij}}{\partial y_i} \right| \right)}.
\end{equation}
Then the partial derivatives of the objective function in each direction are computed. 

Unlike the alternating scheme that separately optimizes positions and dimensions, the non-convex objective function is optimized using the Adam algorithm \cite{kingma2014adam} and to mitigate gradient explosion, gradient clipping is employed. Adam leverages adaptive learning rates and momentum to smoothly traverse flat regions and escape saddle points.

Since $\mathbf{u}_{k+1}$ may violate the hard constraints of the problem \ref{eq_gf}, $\mathbf{u}_{k+1}$ need to be projected onto the feasible region. The projection method is the same as that in PeF \cite{li_pef_2023}. To balance overlap removal and wirelength optimization throughout legalization, the penalty factor is adjusted dynamically according to the wirelength variation between consecutive iterations. Specifically, the expected wirelength variation, denoted by $\Delta \text{HPWL}_{\text{ref}}$, is initialized as:
\begin{equation}
\Delta \text{HPWL}_{\text{ref}} = 0.01 |\text{HPWL}_0|.
\end{equation}

As the optimization progresses and the total overlap area decreases, this reference value $\Delta \text{HPWL}_{\text{ref}}$ is adaptively updated. The penalty factor $\lambda$ is subsequently updated by:
\begin{equation}
\lambda_{k+1} = \min(\lambda_k \mu_k, \lambda_{\text{cap}}),
\end{equation}
where $\lambda_{\text{cap}}$ serves as an upper bound on the penalty factor, and the scaling factor $\mu_k$ is defined as:
\begin{equation}\label{mu}
\mu_k = \mathrm{clip} \left( \mu_0^{ - \frac{\Delta \text{HPWL}_k}{\Delta \text{HPWL}_{\text{ref}}} + 1}, [\mu_{\min}, \mu_{\max}] \right),
\end{equation}
where $\Delta \text{HPWL}_k = \text{HPWL}_{k+1} - \text{HPWL}_k$. In \ref{mu}, $\mu_0$ is set to 1.1, and the scaling factor is constrained within the interval $[0.75, 1.1]$. This adaptive mechanism allows the algorithm to automatically balance the competing objectives of wirelength minimization and overlap reduction throughout the floorplanning process.

\begin{algorithm}[h]
\caption{Global Floorplanning}
\label{alg:jax-glb-algic}
\begin{algorithmic}%[1]
\STATE \textbf{Input:} Netlist \netlist, centers \coord and shapes $\rdim$,
die size $\die$,\\
\hspace*{3.1em}aspect ratio bounds $[\ar_{\min}, \ar_{\max}]$;
\STATE \textbf{Output:} global floorplanning solution $\mathbf{u}(\coord^*,\rdim^*)$.

\STATE $\mathbf{u}_0 \leftarrow (\coord_0,\rdim_0)$

\FOR{$k=0$ to $k_{\max}$}
    \STATE $f_k \leftarrow \lambda_k O(\mathbf{u}_k)
           + \wllse(\mathbf{u}_k)$
    \STATE Compute $\nabla f_k$ 
    \STATE $\mathbf{u}_{k+1} \leftarrow \mathrm{Adam}(\mathbf{u}_k;\nabla f_k)$
    \STATE project $\mathbf{u}_{k+1}$ onto the feasible region;
    \STATE Evaluate $\mathrm{HPWL}_{k+1}$ and real overlap area $O_{k+1}$;
    \STATE Update $\lambda_{k+1}$;
    \IF{$O_{k+1} \le \eta_{\mathrm{olp}}\sum_i w_i h_i$}
        \STATE \textbf{break}
    \ENDIF
\ENDFOR
\STATE \RETURN $\mathbf{u}(\coord^*,\rdim^*)$
\end{algorithmic}
\end{algorithm}

\input{Legalization}

%% file: Legalization.tex
\subsection{Legalization}

Although the global floorplanning provides preliminary physical locations for the modules, the resulting layout still exhibits a small amount of overlap. To remove all module overlaps, legalization is necessary. This section details the algorithms and strategies to eliminate module overlaps.

During legalization, we use the results from global floorplanning as our starting point. Unlike tools such as \cite{li_pef_2023}, which treat soft modules as hard modules and then apply iterative edge-manipulation heuristics in constraint graphs, we instead utilize analytical methods. Specifically, we formulate a hybrid-space convex model that exploits the synergy between linear-domain coordinates and log-domain sizing. This formulation transforms legalization into a tractable convex optimization task, thus guaranteeing global optimality for a given topology.

Based on the results of global floorplanning, we first construct a Horizontal Constraint Graph (HCG) and a Vertical Constraint Graph (VCG) to define the relative positions of the modules.  Instead of comparing the absolute center-to-center distances, we evaluate the signed separations between every pair of modules along both coordinate axes. For two modules $A$ and $B$, the horizontal and vertical signed gaps are defined as
\begin{align}
 \Delta_x
=
|x_A-x_B|
-
\bar d_{AB}^{x}, \nonumber\\
\Delta_y
=
|y_A-y_B|
-
\bar d_{AB}^{y},  
\end{align}
where a positive value indicates that the two modules are already separated along the corresponding axis, while a negative value represents the penetration depth caused by overlap.

\begin{figure}
    \centering
    \includegraphics[width=0.9\linewidth]{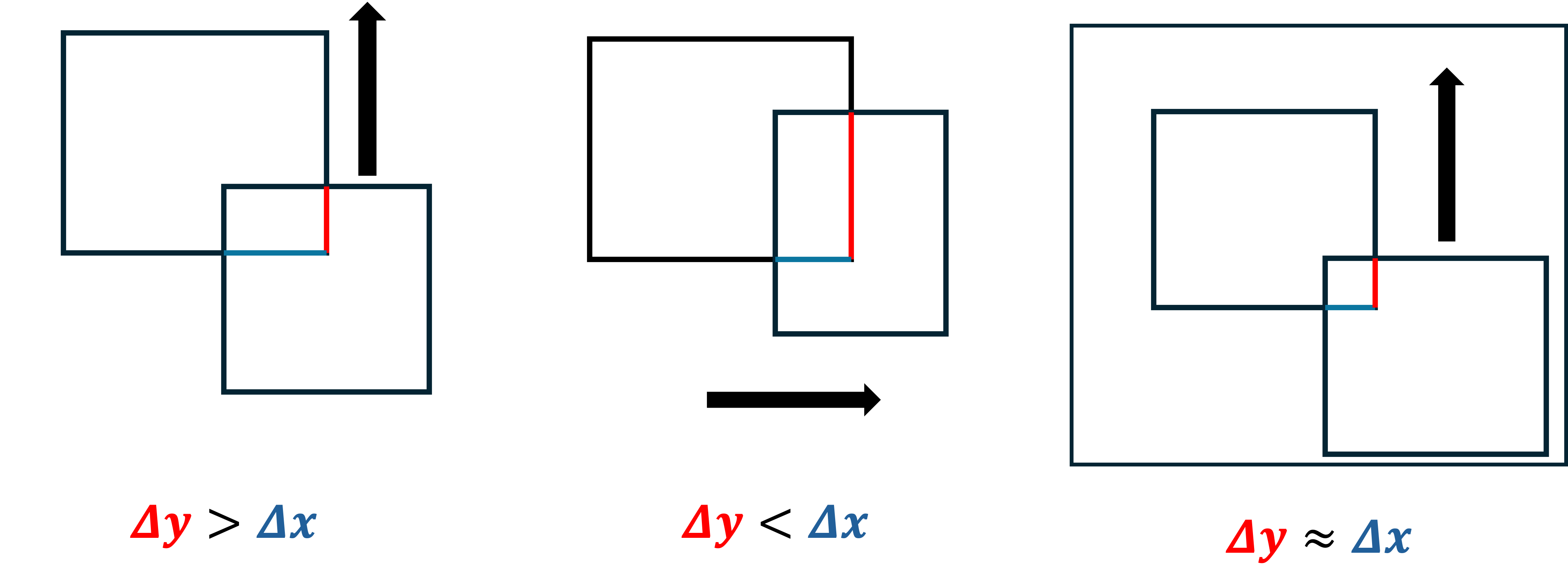}
    \caption{Three relationships between modules which are used to construct VCG and HCG}
    \label{fig:example_HCG_VCG}
\end{figure}

As shown in the Fig. \ref{fig:example_HCG_VCG}, the blue line segments represent the signed gaps in the x-direction, while the red line segments represent the signed gaps in the y-direction. The bold arrows indicate the horizontal or vertical constraints added to the constraint graph. The constraint direction is determined according to the axis requiring the smaller displacement to eliminate overlap. Specifically, if
\begin{equation}
\Delta_x>\Delta_y,
\end{equation}
which corresponds to the minimum translation direction required to separate the two modules. When $\Delta_x=\Delta_y$, the tie is broken according to the aspect ratio of the placement region, favoring the horizontal constraint for wider dies and the vertical constraint for taller dies. Consequently, overlap removal tends to incur the smallest geometric perturbation while preserving the relative placement obtained during global floorplanning.

Once the initial HCG and VCG are constructed, we apply transitive reduction to eliminate redundant edges, thereby streamlining the graph structure and improving computational efficiency. Furthermore, to guarantee a strictly legal layout, we implement a post-process verification step. If any residual overlaps are detected in the resulting floorplan, we identify the specific module pairs lacking explicit constraints, supplement the graphs with the missing edges, and rigorously verify that the updated structures remain valid Directed Acyclic Graphs (DAGs) before finalizing the layout.

While module centers $(x_i, y_i)$ are maintained in the linear domain to preserve the convexity of displacement-based objectives, we map module dimensions into the log-domain by defining $W_i = \ln w_i$ and $H_i = \ln h_i$.  Since dealing with large absolute physical dimensions in the log-domain may easily trigger numerical overflow when calculating exponentials (e.g., $e^{W_i}$), we explicitly normalize all coordinates and dimensions into a $[0, 1]$ scale relative to the maximum die boundary. The proposed legalization model is formulated as follows:
\begin{tcolorbox}[title=Model for initial legalization with convex optimization]
%\footnotesize
\vspace*{-2ex}
\begin{align}\label{eq:wa_gp_obj}
\min\quad & \sum_{e_k \in \mathcal{E}} \wllse(e_k) \nonumber \\ 
& +\beta \sum_{i \in \mathcal{V}} \left( (x_i - x_i^{gf})^2 + (y_i - y_i^{gf})^2 \right) \\
& \hspace*{-3em}\mbox{Area constraints:} \nonumber \\ 
& W_i + H_i \geq \ln A_i, \quad \forall i \in \mathcal{V}\nonumber \\
& \hspace*{-3em}\mbox{Aspect ratio constraints:} \nonumber\\ 
& \ln \ar_{\min} \leq W_i - H_i \leq \ln \ar_{\max}, \quad \forall i \in \mathcal{V} \nonumber\\
& \hspace*{-3em}\mbox{Relative position (HCG, VCG) constraints:} \nonumber \\ 
& x_i + \frac{e^{W_i}+e^{W_j}}{2} \leq x_j, \quad \forall (i,j) \in \text{HCG} \nonumber \\
& y_i + \frac{e^{H_i}+e^{H_j}}{2} \leq y_j, \quad \forall (i,j) \in \text{VCG} \nonumber\\
& \hspace*{-3em}\mbox{Inside die constraints:} \nonumber \\ 
& \frac{e^{W_i}}{2}-x_i \leq 0, \quad \frac{e^{H_i}}{2}-y_i \leq 0, \quad \forall i \in \mathcal{V}\nonumber\\
& x_i + \frac{e^{W_i}}{2} \leq \dwidth, \quad y_i + \frac{e^{H_i}}{2}\leq \dheight,\quad \forall i \in \mathcal{V}. \nonumber
\end{align}
\end{tcolorbox}
To prevent modules from globally drifting and to retain the exact spatial distribution ($x_i^{gf},y_i^{gf}$) of the global floorplanning, we explicitly introduce an $L_2$ displacement penalty term into the objective function \ref{eq:wa_gp_obj}. The proposed legalization model is formulated as a mathematically rigorous convex program by strategically leveraging a hybrid variable space. Through a logarithmic transformation of module dimensions, the traditionally non-linear area and aspect ratio constraints are reduced to linear inequalities, ensuring both convexity and numerical tractability. Meanwhile, the non-overlap and boundary constraints are expressed in the form $f(\mathbf{x}, \mathbf{W}) = \text{affine}(\mathbf{x}) + \text{convex}(\mathbf{W}) \leq 0$; because the exponential function $e^W$ is strictly convex and the sum of convex and affine functions preserves convexity (see~\cite{Boyd2004}, section 3.2.1), these inequalities define a valid convex feasible region. 

It is worth noting that the construction of such a hybrid variable space is not unique. An alternative convexification strategy involves applying an exponential transformation to the coordinate variables $s_{i}= e^{x_i/\alpha}, \quad t_{i}= e^{y_i/\alpha}$ while maintaining module dimensions in the linear domain. By leveraging the monotonicity of the logarithm and the equivalence between minimizing a function $f$ and its logarithm $\log(f)$, the original $\wllse$ can be reformulated into a sum of posynomials. In this alternative framework, the objective function and specific geometric constraints are expressed as posynomials or exponential-based inequalities. Upon applying a global logarithmic mapping, the entire model can be transformed into a standard convex form, often referred to as a Geometric Program (GP) in convex form \cite{boyd2007tutorial}.

% To elucidate the theoretical discrepancy between the log-transformed model \eqref{eq:wa_gp_obj} and the original formulation, we observe that the superiority of the proposed approach stems from a fundamental reconfiguration of the optimization landscape. By ensuring the entire model is Disciplined Convex Programming (DCP)-compliant \cite{grant2008graph}, the transformation guarantees that any locally optimal solution is concurrently the global optimum for a prescribed topology. In contrast, formulating the problem in the original domain introduces significant computational challenges due to the non-convex algebraic expression of the area constraint $g(w, h) = A - wh \leq 0$. Although the two models may appear physically equivalent, they differ fundamentally in computational topology: the original non-convex model necessitates a search over a complex nonlinear manifold, whereas the log-transformed model enables efficient linear partitioning within a stabilized convex space.
\begin{algorithm}%[htbp]
\caption{Legalization (\netlist, \coord, \rdim, \die, \ar)}
\label{alg:gp_wl}

\begin{algorithmic}

\STATE \textbf{Input:} Netlist \netlist, centers \coord and shapes $\rdim$,
die size $\die$,\\
\hspace*{3.1em}aspect ratio bounds \mbox{$\ar$}, factor $\beta$;
\STATE \textbf{Output:} Legalized layout $(\coord^*,\rdim^*)$.
\STATE Construct the HCG and VCG based on the relative positional ordering of centers \coord;

\STATE \textbf{Solve (\ref{eq:wa_gp_obj}):}\\
\hspace*{2em}$(\coord^*,\rdim^*) \leftarrow \text{Solver}(\coord,\rdim)$ with the
logarithmic transformation;

\STATE \textbf{Return:} $(\coord^*,\rdim^*)$.
\end{algorithmic}
\end{algorithm}

Algorithm \ref{alg:gp_wl} details the proposed initial legalization phase. It integrates spatial constraint graph construction with a log-transformed convex optimization framework to optimize module positions and dimensions simultaneously.

%% file: Numericaltests.tex
\section{Experimental Setup and Results}

% The open-source code used in this paper can be found in:
% \textbf{\url{https://github.com/jordicf/CPUPC}}.
% The results of the experiments can be found under the \textit{experiments} folder\footnote{Upon the acceptance of the paper, we plan to upload the experimental results in a public stable repository following the FAIR principles recommended for the best practices of open science. This will incentivate interested researchers to analyze the details of the results and do  comparisons with new tools.}.
% \jordi{The previous paragraph is no longer true. Please, check the footnote also. I think we do not need to make it public at this point. We will make it public when submitted to a journal.}

This section evaluates our proposed fixed-outline floorplanning algorithm using MCNC \cite{mcnc_benchmarks}, GSRC \cite{gsrc_benchmarks}, and HB+ benchmarks \cite{ng2006solving}. The proposed algorithms are implemented in Python. Specifically, the solver in Algorithm~\ref{alg:gp_wl} is IPOPT~\cite{wachter2006implementation}. All experiments were tested on a MacBook Pro equipped with an Apple M5 processor and 16 GB of RAM. We address soft modules floorplanning, using HPWL as the wirelength metric. Comparative data for other state-of-the-art algorithms are obtained directly from their respective publications. 

In our experiments, the hyperparameters for the proposed algorithm were configured as follows: the maximum number of iterations $k_{\max}$ was set to $1000$, the overlap ratio threshold $\eta_{\mathrm{olp}}$ was set to $0.01$ for Algorithm \ref{alg:jax-glb-algic}, and $\beta=10^6$ in convex optimization.

\subsection{MCNC/GSRC Benchmarks}
To evaluate performance, we utilized the \texttt{ami33}, \texttt{ami49}, \texttt{n100}, \texttt{n200}, and \texttt{n300} in the benchmarks, which consist entirely of soft modules with counts corresponding to their names. Our experimental framework comprises two scenarios. The first investigates the impact of the floorplan aspect ratio $\kappa=1$ at a constant 15\% whitespace. The second scenario evaluates performance under a tighter 10\% whitespace constraint with a 1:1 aspect ratio. In both cases, the soft module aspect ratio $[AR_{\min}, AR_{\max}] = [1/3, 3]$. For the first scenario, our algorithm, incorporating global floorplanning and legalization, is compared with the analytical method~\cite{zhan_fixed-floorplanning_2006}, the simulated annealing-based Parquet-4~\cite{adya2004fixed} and the latest state-of-the-art floorplanners SDP~\cite{li_global_2023} and PeF~\cite{li_pef_2023}. To ensure a fair comparison, I/O pad locations are fixed as per the original benchmarks. Results are documented in Table~\ref{tab:benchmarks}. 
\begin{table}[h]
\centering
\caption{HPWL Results of Different Algorithms on MCNC and GSRC Benchmarks(Aspect-ratio $\kappa=1$ and 15\% Whitespace)}
\label{tab:benchmarks}
\begin{tabular}{@{} l rrrrr c @{}}
\toprule
\multirow{2}{*}{\textbf{Algorithm}} & \multicolumn{5}{c}{\textbf{Instances}} & \multirow{2}{*}{\textbf{Ratio}} \\
\cmidrule(lr){2-6}
 & \textbf{ami33} & \textbf{ami49} & \textbf{n100} & \textbf{n200} & \textbf{n300} & \\
\midrule
Parquet-4  & 82149 & 928597 & 342103 & 630014 & 770354 & 1.32 \\
Analytical & 74072 & 799239 & 291628 & 572145 & 702822 & 1.17 \\
SDP        & 65485 & 801402 & 304246 & 553487 & --     & 1.12 \\
PeF        & 64134 & 668608 & 281655 & 509680 & 576158 & 1.02 \\
\textbf{Ours} & \textbf{63224} & \textbf{643721} & \textbf{277459} & \textbf{503778} & \textbf{569338} & \textbf{1.00} \\
\bottomrule
\multicolumn{7}{l}{\footnotesize ``--'' indicates the result is not reported in the literature.}
\end{tabular}
\end{table}

The table reports the HPWL results for each aspect ratio. The final five rows provide the average results of all compared algorithms, normalized to the metrics of our proposed method. Our algorithm consistently achieves the best results across various aspect ratios. Specifically, our approach reduces the average HPWL by at least 2\% compared to PeF, 12\% compared to SDP, 17\% compared to the analytical method, and 32\% compared to Parquet-4.

\begin{table}[htbp]
\centering
\caption{HPWL results of different algorithms on GSRC Benchmarks (Aspect-ratio $\kappa=1$ and 10\% Whitespace)}
\label{tab:scaled_table}
\setlength{\tabcolsep}{1pt} 
\resizebox{\columnwidth}{!}{ 
\begin{tabular}{@{} l *{9}{c} @{}}
\toprule
\textbf{Instances} & \textbf{Parquet-4} & \textbf{Capo 10.2} & \textbf{AR} & \textbf{IMF} & \textbf{IMF+AFF} & \textbf{Ref \cite{lin2018fast}} & \textbf{DeFer} & \textbf{PeF} & \textbf{Ours} \\
\midrule
n100 & 242050 & 224390 & 203700 & 207852 & 208772 & 198649 & 208650 & 194273 & \textbf{189635} \\
n200 & 432882 & 385594 & 367880 & 369888 & 372845 & 351193 & 372546 & 349301 & \textbf{345246} \\
n300 & 647452 & 522968 & 492830 & 489868 & 494480 & 483757 & 498909 & 468235 & \textbf{466365} \\
Ratio & 1.33 & 1.14 & 1.07 & 1.07 & 1.08 & 1.03 & 1.08 & 1.01 & \textbf{1.00} \\
\bottomrule
\end{tabular}
}
\end{table}

In the second experiment, for each benchmark, the aspect ratio was set to $\kappa=1$, and the I/O pads were shifted to the boundary of the floorplanning area. The baseline algorithms for comparison include: analytical methods such as AR \cite{luo_large-scale_2008} and the work by Lin et al. \cite{lin2018fast} (both without a multilevel framework); multilevel-based approaches including Capo 10.2 \cite{adya2004unification}, IMF \cite{chen_new_2008}, and IMF + AMF \cite{lee2003multilevel}; DeFer \cite{yan2010defer}; the simulated annealing-based Parquet-4 \cite{adya2004fixed}; and the electrostatic model-based PeF \cite{li_pef_2023}.

\begin{figure}[H]
    \centering
    \includegraphics[width=0.8\linewidth]{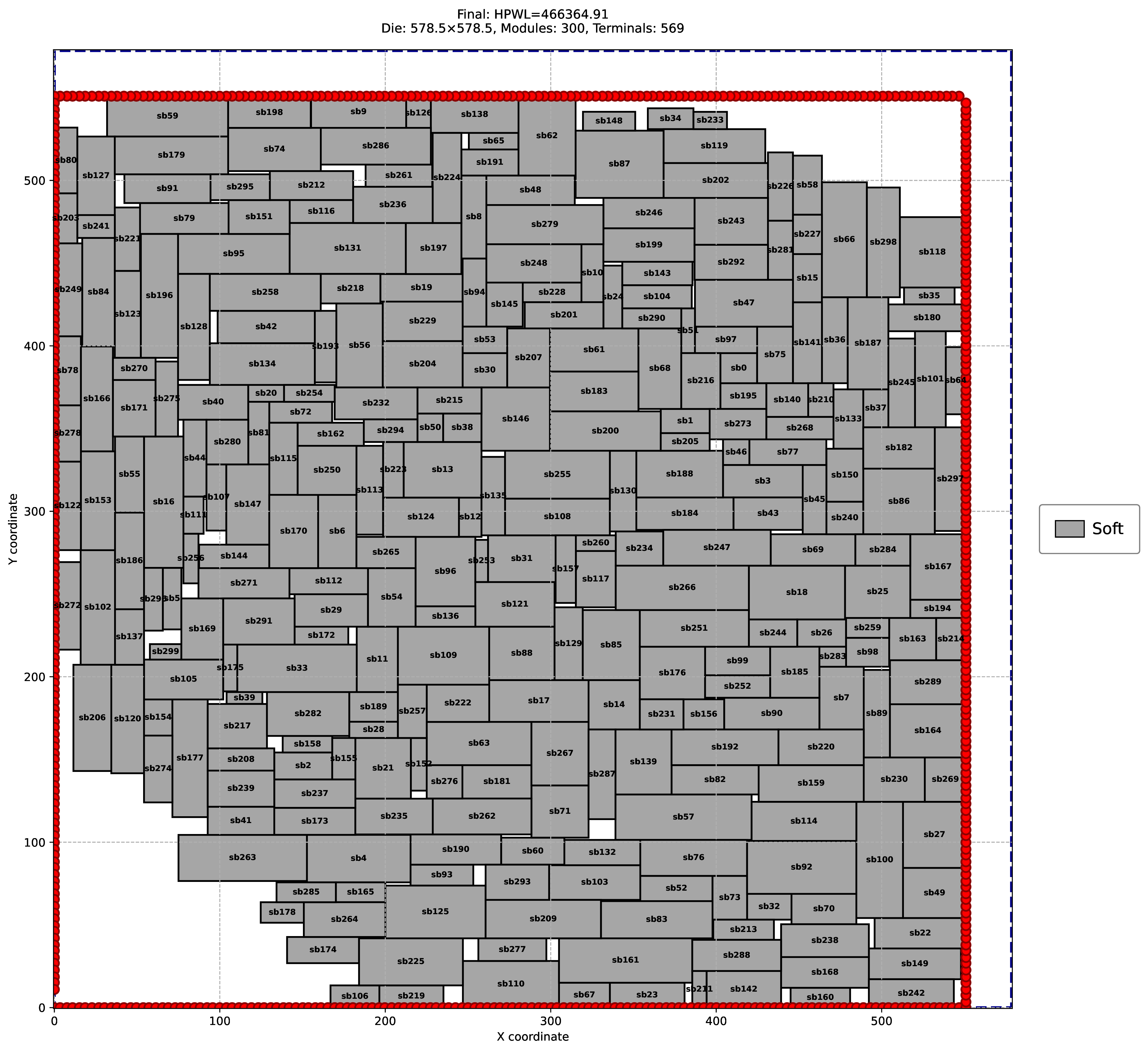}
    \caption{Module packing of \texttt{n300} with 10\% whitespace for fixed-outline floorplanning.}
    \label{fig:n300_0.9}
\end{figure}

The HPWL results for each algorithm are reported in Table~\ref{tab:scaled_table}. All comparative data are obtained directly from \cite{li_pef_2023}. As summarized in the final row of Table \ref{tab:scaled_table}, the experimental results demonstrate that our algorithm consistently achieves the lowest wirelength across all tested GSRC benchmarks. Specifically, our approach outperforms Parquet-4 and Capo 10.2 by 33\% and 14\%, respectively. Furthermore, compared to state-of-the-art techniques such as \cite{lin2018fast} and PeF, our method still provides an average wirelength reduction of approximately 3.0\% and 1.0\%. The floorplanning results of \texttt{n300} are depicted in Fig.~\ref{fig:n300_0.9}.

\subsection{HB+ Benchmarks}
To evaluate the effectiveness of the proposed algorithm, we conduct tests on the HB+ benchmarks which are derived from the original HB benchmark suite by enlarging the largest hard macros by 100\% while proportionally reducing the areas of the remaining soft macros to preserve the total cell area. We compare the proposed method with several representative multilevel and analytical floorplanning algorithms, including DeFer \cite{yan2010defer}, QinFer \cite{ji_quasi-newton-based_2021}, the method in Ref.~\cite{lin2018fast}, and PeF \cite{li_pef_2023}.

\begin{table*}[t]
\centering
\caption{Comparison of HPWL and Runtime on HB+ Benchmarks}
\label{tab:wirelength_time_comparison}
\begin{tabular}{@{} l cccc rrrrr rrrrr @{}}
\toprule
\multirow{2}{*}{Name} & \multirow{2}{*}{Hard} & \multirow{2}{*}{Soft} & \multirow{2}{*}{IO} & \multirow{2}{*}{WS} & \multicolumn{5}{c}{Wirelength ($\times 10^6$)} & \multicolumn{5}{c}{Time (s)} \\
\cmidrule(lr){6-10} \cmidrule(l){11-15}
& & & & & QinFer & Ref.~\cite{lin2018fast} & DeFer & PeF & \textbf{Ours} & QinFer & Ref.~\cite{lin2018fast} & DeFer & PeF & \textbf{Ours} \\
\midrule
HB+01 & 665 & 246 & 246 & 26\% & 2.91 & 3.18 & 3.09 & 3.00 & \textbf{2.88} & 5.0 & 150.0 & 1.8 & 4.3 & \textbf{26.3} \\
HB+02 & 1200 & 271 & 259 & 25\% & 6.47 & 6.80 & 6.17 & 6.02 & \textbf{5.28} & 11.3 & 338.0 & 15.3 & 4.3 & \textbf{61.7} \\
HB+03 & 999 & 290 & 283 & 30\% & 9.17 & 9.68 & 9.19 & 8.17 & \textbf{7.84} & 8.6 & 262.0 & 4.0 & 8.1 & \textbf{47.9} \\
HB+04 & 1289 & 295 & 287 & 25\% & 11.16 & 9.87 & 10.26 & 9.72 & \textbf{8.82} & 12.4 & 209.0 & 14.2 & 7.5 & \textbf{62.0} \\
HB+06 & 571 & 178 & 166 & 25\% & 8.03 & 8.50 & 8.78 & 7.91 & \textbf{7.18} & 7.8 & 125.0 & 5.0 & 7.1 & \textbf{31.6} \\
HB+07 & 829 & 291 & 287 & 25\% & 14.88 & 15.10 & 15.48 & 14.07 & \textbf{13.55} & 8.5 & 230.0 & 4.6 & 7.2 & \textbf{36.1} \\
HB+08 & 968 & 301 & 286 & 26\% & 16.49 & 17.60 & 18.73 & 17.19 & \textbf{15.95} & 22.8 & 324.0 & 19.3 & 9.7 & \textbf{52.8} \\
HB+09 & 860 & 253 & 285 & 25\% & 17.24 & 18.30 & 16.66 & 15.81 & \textbf{14.67} & 8.7 & 259.0 & 4.2 & 9.0 & \textbf{41.9} \\
HB+10 & 809 & 786 & 744 & 20\% & 42.33 & 46.70 & 45.12 & 40.61 & \textbf{39.58} & 11.4 & 191.0 & 6.3 & 32.9 & \textbf{77.0} \\
HB+11 & 1124 & 373 & 406 & 25\% & 25.70 & 28.20 & 26.99 & 24.58 & \textbf{24.13} & 10.1 & 494.0 & 7.1 & 9.3 & \textbf{58.1} \\
HB+12 & 582 & 651 & 637 & 26\% & 52.83 & 53.60 & 50.17 & 48.96 & \textbf{46.98} & 16.6 & 136.0 & 5.5 & 15.0 & \textbf{55.9} \\
HB+13 & 830 & 424 & 490 & 25\% & 35.44 & 35.40 & 35.51 & 32.65 & \textbf{31.45} & 23.7 & 90.0 & 5.9 & 14.2 & \textbf{38.4} \\
HB+14 & 1021 & 614 & 517 & 25\% & 60.68 & 63.40 & 64.50 & 59.62 & \textbf{58.55} & 14.9 & 208.0 & 12.0 & 16.7 & \textbf{81.9} \\
HB+15 & 1019 & 393 & 383 & 25\% & 77.48 & 79.10 & 84.29 & 73.93 & \textbf{72.28} & 37.1 & 532.0 & 14.7 & 24.8 & \textbf{87.2} \\
HB+16 & 633 & 458 & 504 & 25\% & 98.53 & 92.90 & 98.66 & 87.32 & \textbf{86.53} & 11.2 & 160.0 & 8.1 & 20.0 & \textbf{81.2} \\
HB+17 & 682 & 760 & 743 & 25\% & 140.84 & 140.00 & 144.56 & 138.44 & \textbf{132.43} & 15.2 & 169.0 & 14.7 & 20.5 & \textbf{107.3} \\
HB+18 & 658 & 285 & 272 & 25\% & 70.36 & 70.70 & 71.86 & 68.05 & \textbf{65.95} & 12.4 & 66.0 & 11.3 & 14.5 & \textbf{81.1} \\
\midrule
Ratio & \multicolumn{4}{c}{---} & 1.05 & 1.08 & 1.08 & 1.00 & \textbf{0.95} & 1.23 & 24.43 & 0.89 & 1.00 & \textbf{5.4} \\
\bottomrule
\end{tabular}
\end{table*}

Table \ref{tab:wirelength_time_comparison} summarizes the characteristics of the HB+ benchmark suite together with the experimental results of the compared floorplanning algorithms. Other benchmark specifications, such as the floorplan dimensions and the upper and lower aspect-ratio bounds of the soft macros, are identical to those reported in the original benchmark descriptions and are therefore omitted here for brevity.

As shown in the last row of Table \ref{tab:wirelength_time_comparison}, the proposed method achieves the smallest average normalized HPWL, reducing the average wirelength by 10\%, 13\%, and 13\% compared with QinFer, DeFer, Ref.~\cite{lin2018fast}, respectively. In addition, our method consistently outperforms PeF, yielding an average HPWL reduction of approximately 5\%.
 Columns 11--15 of Table~\ref{tab:wirelength_time_comparison} compare the runtimes of the evaluated algorithms. As shown in the last row, the proposed method requires approximately $5.01\times$ the runtime of PeF, which is the fastest analytical floorplanning algorithm among the compared methods. This additional computational cost mainly arises from the iterative gradient-based optimization. Nevertheless, the runtime remains within a practical range for all HB+ benchmarks requiring only 26.3 to 107.3 seconds. More importantly, the increased runtime is accompanied by a consistent improvement in solution quality.

 It is worth noting that both PeF and the method in Ref.~\cite{lin2018fast} are adapted from analytical placement algorithms. For the relatively small benchmarks in Table \ref{tab:scaled_table}, neither method employs a multilevel framework. Under this setting, the proposed method achieves an average HPWL reduction of 1--3\% compared with PeF and Ref.~\cite{lin2018fast}. For the larger HB+ benchmarks in Table \ref{tab:wirelength_time_comparison}, however, Ref.~\cite{lin2018fast} adopts a multilevel optimization framework, whereas both PeF and the proposed method remain flat optimization approaches without multilevel decomposition. Despite not using a multilevel framework, the proposed method further increases its advantage over Ref.~\cite{lin2018fast}, reducing the average HPWL by 11\% while requiring only about one-quarter of its runtime. This result suggests that the proposed continuous optimization framework scales effectively to larger floorplanning instances and can maintain high solution quality without relying on multilevel techniques. The increasing performance gap on larger benchmarks also demonstrates the robustness and scalability of the proposed flat optimization strategy.

% \subsection{FloorSet Benchmarks}

% FloorSet is a comprehensive, open-source dataset and benchmark designed to train and evaluate Machine Learning (ML) approaches for VLSI floorplanning. Rather than collecting layouts from existing designs, FloorSet generates near-optimal floorplans by statistically modeling the characteristics of modern industrial SoC designs.

% To further evaluate, we select the 10 largest benchmarks from the FloorSet-Lite test dataset for testing. These benchmarks are not included in the training set and are accompanied by near-optimal reference ("golden") floorplans provided by FloorSet.

\subsection{Runtime Analysis}
This section provides a detailed analysis of the computational efficiency of the proposed floorplanning framework. We evaluate the runtime performance based on the data presented in Table \ref{tab:benchmarks} and Table \ref{tab:wirelength_time_comparison}, where the total runtime is the sum of the runtime of the three stages of the flow.
\begin{figure}%[H]
    \centering
    \includegraphics[width=0.8\linewidth]{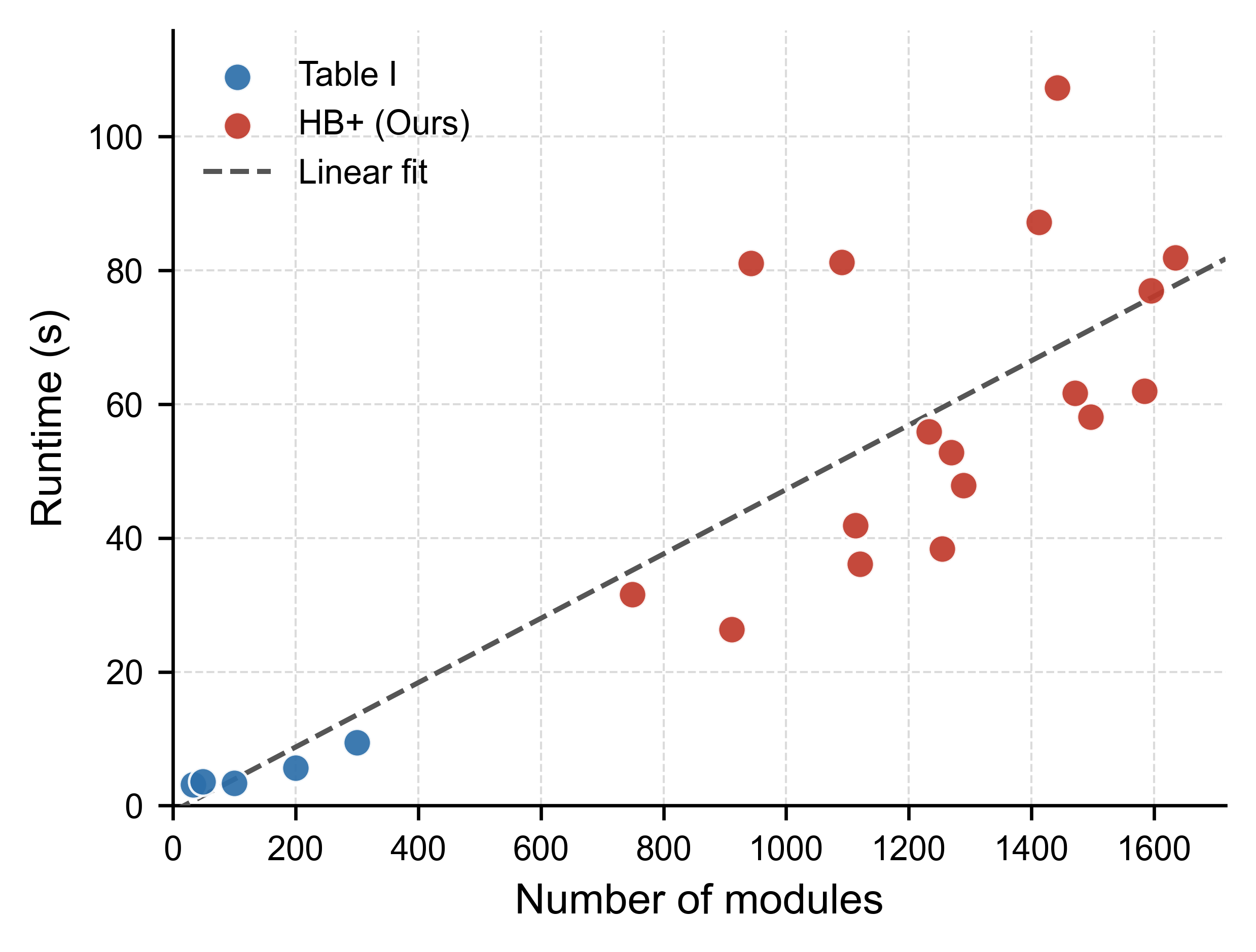}
    \caption{Total runtime on GSRC, MCNC and HB+ benchmarks}
    \label{fig:runtime}
\end{figure}

In the Fig. \ref{fig:runtime}, the dashed line are the liner
fittings of the average run-times of our floorplanning algorithm. Since the total runtime of the proposed global floorplanning algorithm is approximately equal to the number of iterations multiplied by the average runtime per iteration, we evaluate the efficiency in terms of the average runtime of a single iteration. As the maximum number of iterations is fixed at 1000, the average runtime per iteration remains on the order of milliseconds even for the HB+ benchmarks, demonstrating the high computational efficiency of the proposed method and matching our expectations.

%% file: Conclusion.tex
\section{Conclusion}

This article presents a unified analytical framework for VLSI global floorplanning and legalization. For global floorplanning, we propose a continuously differentiable objective with an adaptive, Adam-based projected gradient method. This eliminates piecewise geometric enumeration, making it highly amenable to hardware acceleration. To eliminate residual overlaps, we introduce a DCP-compliant, hybrid-space convex legalization model that linearizes non-convex area constraints using logarithmic mapping and constraint graphs (HCG/VCG). Ultimately, this synergistic approach delivers a theoretically rigorous, efficient, and scalable solution to modern floorplanning challenges.

Extensive experiments on MCNC, GSRC, and HB+ benchmarks validate the scalability and superiority of our flat optimization approach. In MCNC/GSRC experiments, we achieves the best HPWL, outperforming simulated annealing methods (Parquet-4) by up to 33\% and state-of-the-art analytical tools (e.g., PeF, SDP) by 1\% to 12\% under strict 10\% and 15\% whitespace constraints. In HB+ benchmarks, we demonstrates robust scalability on larger designs without relying on multilevel framework decomposition. It yields a 5\% to 13\% HPWL reduction over recent state-of-the-art floorplanners while maintaining highly practical runtimes. In future work, we plan to improve the framework to handle large-scale industrial designs like FloorSet \cite{mallappa2024floorset} and extend the methodology to non-rectangular module instances, making the algorithm valuable for both academic and industrial floorplanning applications.

%% file: Acknowledgment.tex
\subsection*{\textbf{Acknowledgements}}
The authors would like to thank Jalal Benallal, Mohammed Wahbi, and Nandyala Guptha Vivekanand, from Qualcomm (Cork, Ireland), for their valuable assessment and insightful discussions.

%% file: Appendix.tex
% \appendix
\appendices